# Quantum computing with spin qubits in semiconductor structures

Vladimir Privman, Dima Mozyrsky, Israel D. Vagner

Department of Physics and Center for Advanced Materials Processing,
Clarkson University, Potsdam, New York 13699-5820, USA

Electronic address: privman@clarkson.edu

**Abstract**

We survey recent work on designing and evaluating quantum computing implementations based on nuclear or bound-electron spins in semiconductor heterostructures at low temperatures and in high magnetic fields. General overview is followed by a summary of results of our theoretical calculations of decoherence time scales and spin-spin interactions. The latter were carried out for systems for which the two-dimensional electron gas provides the dominant carrier for spin dynamics via exchange of spin-excitons in the integer quantum Hall regime.

PACS: 73.20.Dx, 71.70.Ej, 03.67.Lx, 76.60.-k

Keywords: quantum computing, semiconductor, decoherence, spin, qubit, electron gas

## 1. Introduction

The field of quantum computing has seem explosive growth of experimental and theoretical interest. The promise of quantum computing [1-5] has been in exponential speedup of certain calculations via quantum parallelism. In Figure 1, the top flow chart shows the "classical" computation which starts from binary input states and results in binary output states. The actual dynamics is not really that of Newtonian classical mechanics. Rather the computation involves many-body irreversible "gate" device components, made of semiconductor materials in modern computers, which evolve irreversibly, "thermodynamically" according to the laws of statistical mechanics. As the size of the modern computer components approaches atomic, the many-body quantum behavior will have to be accounted for in any case [6].

The idea of quantum computing, however, is not just to account for, but to actually utilize the quantum-mechanical dynamical behavior. This is not an easy task. Quantum mechanics allows for parallelism in evolution: one can "process" a linear superposition of several input states at once, as illustrated in the lower flow chart in Figure 1. The price paid is that coherent processing of information, according to the law of quantum mechanics, must be accomplished in systems much larger than atomic-size (or more importantly, with many degrees of freedom). There are numerous conceptual and experimental obstacles to accomplishing this task, that have generated a lot of interest, excitement, and new results in computer science, physics, and engineering.

The functioning of a quantum computer involves initialization of the input state, then the actual dynamical evolution corresponding to computation, and finally reading off the result. Various specific requirements for implementation have been identified [2-5]; here we provide only a limited introductory overview.

Let us begin by considering the reading off of the final result. The reason for the question mark in the lower chart in Figure 1 is that quantum measurement of the final superposition state can erase the gain of the parallel dynamics, by collapsing the wave function. Therefore, a key issue in quantum computing has been to find those algorithms for which the readout of the final state, by way of projecting out a certain average property, still retains the power of the quantum parallelism. To date, only few such examples are known [1,3,4,7], the most celebrated being the Shor algorithm [1] for factoring of integers, the invention of which boosted quantum computing from an obscure theoretical field to a mainstream research topic.

The preparation of the initial state does not seem to present a problem for most quantum computing realizations [2-5], accept perhaps the ensemble liquid state NMR approach [8,9] which relies on the initial thermal distribution to produce deviation of the density matrix from the equal-probability mixture state. In most other approaches, the initial state can be produced by first fully polarizing the quantum bits (qubits), i.e., putting them in one of the two quantum levels. Note that we consider two-state qubits here, realized, for instance, by spins 1/2 of nuclei or gate- or impurity-bound electrons, in applied magnetic field. The fully polarized state is then subject to gate operations to form the desired input state. Part of a quantum-computing algorithm should be the prescription on how to choose the initial state to represent the classical information of the input, like the input integer in the factoring. In most cases, this prescription is easily accomplished by single-qubit and two-qubit gates.

The actual dynamical evolution (the process of computation) in quantum computing is fully reversible and nondissipative, unlike classical computing. Much progress has been made in resolving both the conceptual and computer-engineering "design" issues for quantum computation. Specifically, the computation can be carried out [2-5,10-13] by a universal set of gates: single-qubit rotations and nearly any two-qubit gate. The gates are not connected in space like in classical computers but are activated in succession in time, to control single-spin dynamics and also switch on and off two-spin interactions (we use "spin" and "qubit" interchangeably).

Many interesting matters have been resolved, which are not reviewed here. These include the understanding of how the finiteness of the state space (i.e., two states for spin 1/2) replaces the "classical" digitalization in quantum computing. Also, the "classical" copying (fan-out) function is not possible in quantum mechanics. It is replaced by entanglement with ancillary qubits to accomplish redundancy needed for error correction [14-20]. Sources of errors due to interactions with environment in quantum mechanics involve not only the usual relaxation (thermalization) but also loss of coherence [21-28]. This quantum decoherence (dephasing) can be faster than relaxation because it does not require energy exchange.

A conceptually important issue has been the scalability of quantum computing: can one process macroscopically large amounts of information by utilizing quantum error correction based on redundancy via entanglement with ancillary qubits? The affirmative answer to this question has been one of the triumphs of the theory [14-20]. It provided a new paradigm for emergence of controlled/organized macroscopic behavior from microscopic dynamics, on par with the conceptual possibility of living organisms, which we observe by cannot yet "manufacture," and million-gate classical computers which are man-made.

With all these theoretical advances at hand, the next step is to ask whether a man-made quantum computer can be realized? There have been several experimental directions of exploration, most presently are still at the level of one or two qubits, or, for ensemble liquid-state NMR, which emulates quantum dynamics by evolution of the density matrix of a large collection of molecules, 5-7 qubits.

In this introductory survey, we summarize results of our work on two-spin interactions and spin decoherence in semiconductor heterostructures. In Section 2, we consider the spin-based quantum computing proposals in such systems. Time scales of relaxation and decoherence are addressed in Section 3. Finally, Section 4 reports results for models with nuclear spins as qubits.

**2. Spin-Based Quantum Computing in Semiconductor Heterostructures**

The general layout of a solid-state quantum computer is shown in Figure 2. Qubits are positioned with precision of few nanometers in a heterostructure. One must propose how to effect and control single-qubit interactions, two-qubit interactions, and explore how the controlled dynamics owing to these interactions compares to decoherence and relaxation. The proposal must include ideas for implementation of initialization, readout, and gate functions.

The first proposal including all these components was for qubits realized in an array of quantum dots [29] coupled by electron tunneling. The first spin-based proposal [30] utilized nuclear spins coupled by the two-dimensional electron gas, the latter in the dissipationless integer quantum Hall state [31] that requires low temperatures and high magnetic fields. An important advancement was the work of Kane [32] where gate control of nuclear-spins of donor impurities, separated less than 10 nm and coupled via the outer impurity electrons which are bound at low temperatures, was proposed. Most of these ideas also apply to electron-spin qubits, bound at impurities, in quantum dots, or directly by gates. Several elaborate solid-state heterostructure quantum computing schemes have been proposed in the literature recently [28,33-41]. There are also other promising proposals involving surface geometries: superconducting electronics [42-46] and electrons on the surface of liquid helium [47].

There have been several planned and ongoing experimental efforts [32-36,43-45,48-54] ultimately aimed at solid-state quantum computing and other quantum information processing realizations. The final geometry is expected to be most sensitive to the implementation of readout, because it involves quantum measurement, i.e., supposedly interaction with or transfer of information to a macroscopic device. Therefore, much of the experimental effort presently has been focused on single-qubit (single-spin) measurement approaches.

The theoretical efforts can be divided into two majors tasks. The process of single-spin measurement must be understood for the readout stage of quantum computing. Several conceptual and calculations advancements have been made in understanding quantum measurement [26,32-36,46,50,55,56] as it applies to atomic-size qubit systems interacting with environment and typically "measured" directly by the effect of the spin-qubit state on transport, or first transferring the spin state to a charge state that is easier to measure, e.g., in single-electron transistors and similar devices.

In this survey, we outline results of the second evaluation task: that of understanding the processes and times scales involved in the dynamics of the actual computation. As summarized in Figure 3, this main stage of the quantum computation process involves control of spins and their interactions. It also involves processes that we do not control and are trying to minimize: relaxation and decoherence.

Control of individual qubits is usually accomplished externally. For nuclear spins, NMR radio-frequency radiation can be used, see Figure 2. For electron spins, the ESR microwave frequencies are suitable. Such radiation cannot be focused on the scale of 10-100 nm. Instead, selectivity must be accomplished by independent means. Several proposals exist, the most promising being control by gates. The applied gate voltage modifies the electronic wave function changing interactions and therefore resonant frequencies. We will denote the time scale of the external single-qubit control by $T_{ext}$. This can be the Rabi time of a spin flip.

The qubit-qubit interactions are typically assumed to be mediated by electrons that "visit" both qubit environments. For instance, in liquid-state ensemble NMR [8,9] with complex molecules, or in the original model [32] of phosphorous impurity donors in silicon, the wave functions of the valence, outer electrons of nearby qubits overlap. Specifically, in the P donor case, the single outer electron of the donor atom remains bound at low temperatures but has orbital radius of order 2 nm owing to the large dielectric constant of the silicon host. Therefore, it is hoped that these electrons, in nearby donors positioned as in Figure 2, will mediate nuclear-spin qubit interactions.

Our approach [27,28] allows for larger qubit separation, up to order 100 nm, by relying on the two-dimensional electron gas in the heterostructure to mediate qubit-qubit interactions. This two-dimensional electron gas is usually obtained by spontaneous or gate-induced transfer of electrons from impurities to the two-dimensional interface layer in which the qubits are positioned. The source impurities are located at some separation from this layer or in the bulk. The two-dimensional electron gas can be made nondissipative in certain ranges of large applied magnetic fields at low temperatures, when these conduction electrons in the layer are in the integer quantum Hall effect state. Owing to this property and also larger qubit separation allowed, we consider this the most promising approach and focus our present review on such systems.

The time scale of the qubit-qubit interactions will be devoted by $T_{int}$. This is the time it takes to accomplish a two-qubit quantum gate, such as CNOT [2-5,57]. Typically for semiconductor quantum computing proposals, $T_{int} < T_{ext}$, and in fact the case with $T_{int} \ll T_{ext}$ has some advantages because one can use several fast single-spin flips to effectively switch interactions of some qubits off over the gate cycle. Another approach to controlling (on/off) of the two-qubit interactions is by gates, see Figure 2, which affect the two-dimensional electron gas and the localized electron wavefunctions.

However, the same conduction electrons that provide the qubit-qubit interactions, also expose the qubits to the environment, causing relaxation and decoherence. Other interactions will also be present, that play no role in the useful quantum-computing dynamics but contribute to these undesirable processes. Relaxation and decoherence, and their associated time scales, are addressed in the next section.

## 3. Time scales of relaxation and decoherence

The processes of relaxation and decoherence considered here [21-28] are associated with the dynamics of a small, few-qubit quantum system as it interacts with the environment. Ultimately, for a large, multi-qubit system, many-body quantum chaos-like behavior must also be accounted for, and some advances in model system studies have been reported recently [5,58]. Our discussion here will be for the few-qubit case mostly because it allows more system-specific investigations for actual quantum-computing proposals.

Dynamical processes that are unwanted in quantum computing, because they result from the environmental influences rather than from the controlled radiation pulses and gate potentials, can proceed on various time scales. In fact, it is not guaranteed that processes of various types, relaxation/thermalization vs. decoherence/dephasing, can even be unambiguously distinctly identified.

At low temperatures, it is generally hoped that thermalization, which requires transfer of energy, slows down. If the fastest such processes proceed on times scales of order $T_1$, then this time increases at low temperatures because there are less excitations (phonons, electron gas modes, etc.) to couple the small quantum system to the rest of the solid-state host material.

On the other hand, processes that do not require flow of energy to or from the environment, can still effect the phase of the quantum-superposition amplitudes and cause decoherence. These processes can thus proceed faster, on the time scale $T_2$. While these comments seem to suggest that $T_2 \leq T_1$, there is no obvious reason to have generally $T_2 \ll T_1$ at low temperatures.

However, if the spectrum of the dominant excitations mediating the qubit coupling (both to each other and to the host material) has a gap, then we expect that all the relaxation and decoherence processes will be suppressed. Furthermore, the suppression of the relaxation will be exponential, with the Boltzmann factor for that energy gap. Then, $T_2 \ll T_1$ will be satisfied but also, more importantly, the actual values of both time scales will be inordinately large. This was found, theoretically and experimentally, to be the case for the integer-quantum-Hall-state two-dimensional electron gas as mediator of the localized-spin (nuclear, electronic) coupling in semiconductor heterostructures [27,28,59-63].

It is important to emphasize that relaxation and decoherence are really many-body properties of the system plus environment. Entanglement with the environment owing to the unwanted couplings results in the small quantum system having no pure wavefunction even if initially it was prepared in a pure state. Instead, it can be described by a statistical mixture represented by a density matrix, once the environment is traced over.

This reduced density matrix of the system is expected to evolve to the thermal one at large times. The approach to the thermal density matrix, which is diagonal in the system-energy basis, defines the time scale $T_1$. If the temperature is low enough, then there is the expectation, see [25,26] and references therein, that for some intermediate time scales, of order $T_2$, the density matrix becomes nearly-diagonal in a basis which is determined not by the systems Hamiltonian (energy), but by the interaction operator with the environment. This latter process corresponds to loss of quantum coherence.

As emphasized in Figure 3, evaluation of a quantum-computing proposal requires, among other things, establishing the relation $T_{ext}, T_{int} \ll T_2, T_1$. Owing to calculational difficulties, the single-qubit times $T_{1,2}$ will usually be used, though, as mentioned earlier, some study of the multi-qubit "quantum chaos" effects may be required. For spin-qubit quantum computing in semiconductor heterostructures, the relation is typically $T_{ext} \ll T_{int} \leq T_2 \ll T_1$, so the issue is usually how small is the quality ratio $Q = T_{int} / T_2$.

The required value of $Q$, needed for fault-tolerant quantum error correction, depends on the physical model of error sources and can be as small as $Q = 10^{-6} - 10^{-4}$, see [15,18-20], or as large as $Q = 1/2$, see [64]. For the systems of interest to us here, spin qubits in

– 6 –

semiconductor structures, the value of $Q = 10^{-5}$ is a reasonable working estimate. Thus, we seek systems/conditions with $T_{int}/T_2 \leq 10^{-5}$.

**4. Results for nuclear-spin qubits**

In this section we outline results for models of quantum computing with nuclear spins as qubits, and with coupling mediated by the two-dimensional electron gas in the integer quantum Hall effect state [27,28,30]. In strong magnetic fields, the spatial states of the electrons confined in the two-dimensional layer in which the qubits are placed, see Figure 2, are quantized by the field to resemble free-space Landau levels. The lattice potential and the impurities actually cause formation of narrow bands instead of the sharp levels, separated by localized states. As a result, for ranges of magnetic field, the localized states fill up while the extended states resemble completely filled integer number of Landau levels. These states are further Zeeman split owing to the electron spin. At low temperatures, one can find field values such that only one Zeeman sublevel is completely filled in the ground state.

The electronic state in such systems, that show the quantum Hall effect [31] in conductivity, are highly correlated and nondissipative. If nuclear spins are used as qubits, i.e., atoms with nuclear spin 1/2 are sparsely positioned in the zero-nuclear spin host, such as the zero-nuclear-spin isotope 28 of Si, which constitutes 92% of natural silicone, then their zero-temperature relaxation will be significantly slowed down: experimentally, $T_1 \simeq 10^3$ sec [62].

Localized spins, both nuclear and electronic, interact by exchanges of spin excitons—spin waves consisting of a superposition of bound electron-hole pair states. The spectrum of these excitations [65,66], observed experimentally in [67], has a gap corresponding to the Zeeman splitting. This gap is the cause of slow relaxation and decoherence. The exchange of virtual spin excitons mediates the qubit-qubit interaction and also, via scattering of virtual excitons from impurity potentials, relaxation and decoherence of single qubits.

The original proposal to use nuclear spin qubits directly coupled by the two-dimensional electron gas [30], required positioning the qubits at distances comparable to several magnetic lengths. The latter is of order 10 nm for magnetic fields of several Tesla. The qubit-qubit interaction decays exponentially on this length scale. Recently, we proposed a new improved model [28] in which the qubit interactions are mediated via coupling of the two-dimensional electron gas to the outer impurity electrons. This applies if the atoms, whose nuclear spins are the qubits, are single-electron donors such as the isotope 31 of P. These phosphorous impurities were originally utilized in the model of Kane [32] where they must be actually positioned at separations of about 4 nm for the wavefunctions of the outer electrons, which are bound at low temperatures, to overlap significantly.

In our new improved model [28], with nuclear spins coupling to the outer bound electrons which, in turn, interact via the two-dimensional electron gas, the interaction turned out to be of a much longer range as compared to the model of [32]: the qubit separation can be of order 100 nm. Another advantage is that gate control of the individual qubits and of qubit-qubit interactions is possible. We have carried out extensive perturbative many-body calculations [27,28,30,68] allowing estimation of $T_{int}$ and $T_2$ for both the original quantum-computing proposal [30] and its improved version [28], where the main improvement is in the possibility of the gate control along the lines of [32]. The "clock speed" of the improved model is also faster by about two orders of magnitude. The technical details of these rather cumbersome calculations are available in the literature and will not be detailed here.

The results are summarized in Table 1. We show estimates of all four relevant time scales for the two models introduced earlier. The "original" model [30] corresponds to nuclear spins 1/2 introduced at qubits in atoms without an outer loosely bound electron. The "improved" model corresponds to the case when the outer electron is present and its interaction with the nuclear spin and the two-dimensional electron gas dominates the dynamics.

The data shown in Table 1 were obtained assuming typical parameters for the standard heterojunctions utilized in quantum-Hall-effect experiments today, and qubit separation of 65 nm. Thus, the parameter values taken [28,30] were more appropriate for the GaAs system than for Si, even though the main isotopes of gallium and arsenic have nuclear spin 3/2 and cannot serve as spin-zero hosts. The reason for using these values has been that experimental verification of some of the numbers might be possible in the available materials before cleaner and different composition materials needed for quantum computing are produced.

Our estimates, see Table 1, indicate that the quality factor $Q = 10^{-5}$ is not obtained for the present system. Actually, no quantum computing proposal to date, scalable by other criteria, satisfies the $10^{-5}$ quality-factor criterion. The values range from $10^{-1}$ to $10^{-2}$. The resolution could come from development of better error-correction algorithms or from improving the physical system to obtain a better quality factor. In our estimation of the decoherence time scale, we used parameters typical of a standard, "dirty" heterostructure with large spatial fluctuations of the impurity potential. These heterostructures have been suitable for standard experiments because they provide wider quantum-Hall plateaus, i.e., ranges of magnetic field for which all the extended states of a Zeeman sublevel are filled. Much cleaner, ultra-high mobility structures can be obtained by placing the ionized impurity layer at a larger distance from the two-dimensional gas or by injecting conduction electrons into the heterostructure by other means. Thus, our quantum-computing proposals [28,30] are unique not only in the large qubit separation allowed but also in that there is a clear direction of exploration to allow physical, rather than algorithmic, resolution of the quality factor problem. This possibility should be further explored both experimentally and theoretically.

The authors acknowledge useful discussions and collaboration with M. L. Glasser, R. G. Mani and L. S. Schulman. This research was supported by the National Security Agency (NSA) and Advanced Research and Development Activity (ARDA) under Army Research Office (ARO) contract number DAAD 19-99-1-0342.

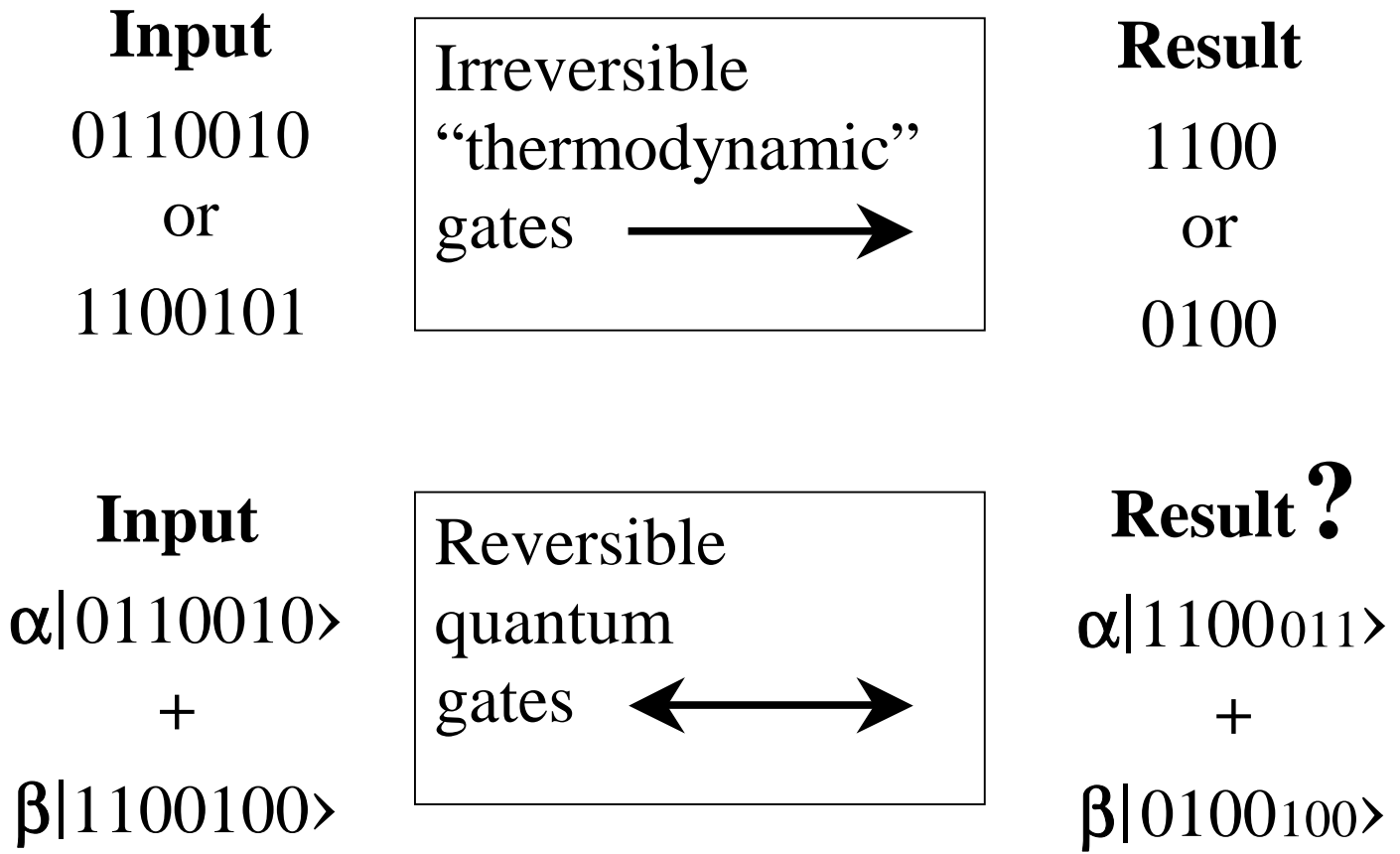

**Figure 1.** Comparison of the classical and quantum approaches to computing. The upper flow chart schematically represents implementation of a traditional irreversible "classical" computation process, where transformation of the input set of bits into the result is accompanied by a succession of irreversible gates. Owing to their irreversibility, the gates can be connected in space rather than switched on and off at different times. The lower flow chart shows quantum processing of information, where the input and the final result are both in superposition states, yielding quantum parallelism. The dynamics is reversible: there is a one-to-one correspondence between the initial and final states. Therefore, number of the input and output quantum bit (qubits) is the same even though some of the output qubits (set in a smaller font) might not be used in the final extraction of the classical result by measurement. The quantum gates are applied in succession by being switched on and of at different times during the computation. The question mark signifies the difficulty of finding quantum algorithms that retain the power of quantum parallelism after measurement needed to read off the final result as classical information.

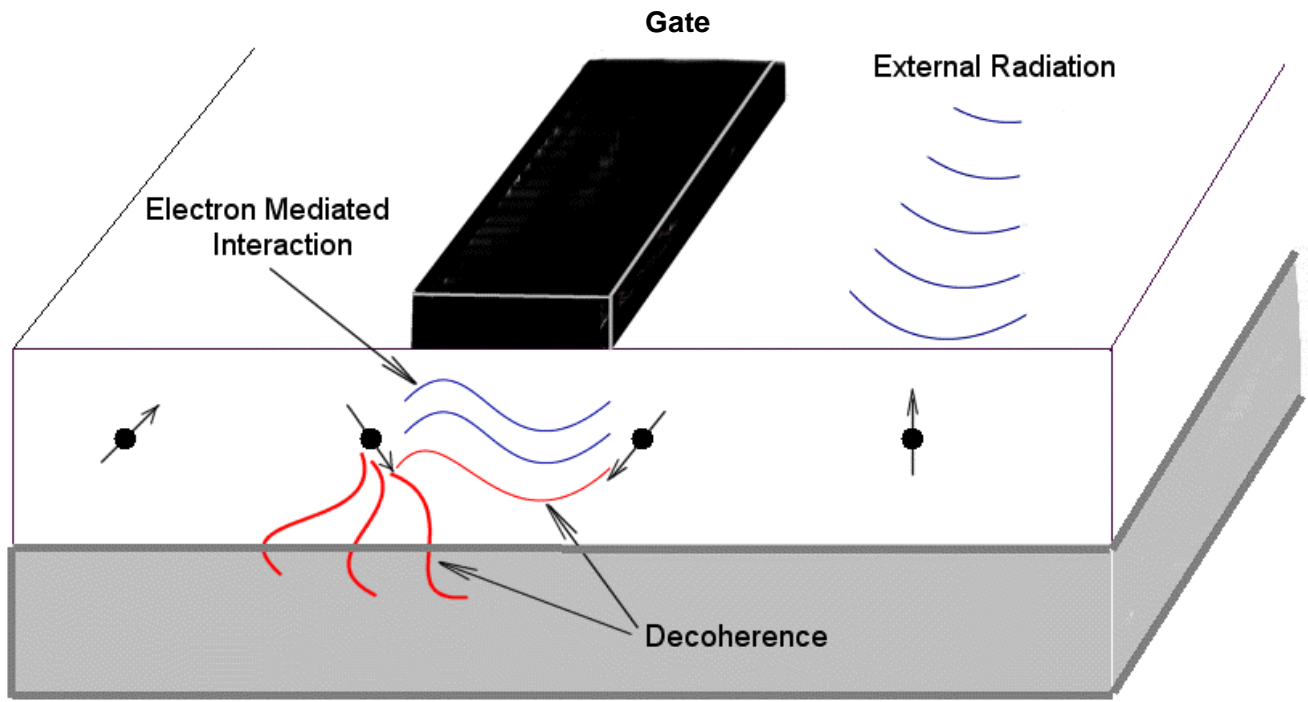

**Figure 2.** Schematic illustration of a semiconductor heterostructure quantum information processor. The qubits, represented by the arrows overlaying heavy dots, are spins 1/2 of nuclei or localized electrons. Individual control of the temporal evolution of the spins can be achieved with the use of external electromagnetic radiation, i.e., NMR or ESR pulses. The spins are also coupled with each other via interaction mediated by the two-dimensional electron gas in the heterostructure, or by other means. The external and internal interactions can be controlled by gates formed on top of the heterostructure. The external environment, that includes crystal lattice, electron gas, defects, impurity potentials, causes relaxation and decoherence of the qubits.

**Initialize**

**Control and
evolve in time**
$T_{ext}, T_{int} \ll T_2, T_1$
$\left\{\begin{array}{ll} \textbf{Control qubits:} & T_{ext} \\ \textbf{\& interactions:} & T_{int} \;(> T_{ext}) \\ \\ \textbf{Avoid relaxation:} & T_1 \\ \textbf{\& decoherence:} & T_2 \;(\leq T_1) \end{array}\right.$

**Measure**

**Figure 3.** Evaluation of quantum computing models. One of the criteria for feasibility of quantum computing in a given physical system is the possibility of initialization of the qubits in the desired superposition state. Another important design consideration is control of qubit states and of their interactions. In order to implement quantum computing effectively, the time scales for realization of single and two-qubit logic gates, $T_{ext}$ and $T_{int}$, respectively, should be several orders of magnitude smaller than the time scales of relaxation and decoherence, $T_1$ and $T_2$. The relationships between these time scales are further explained in the text. Finally, efficient and reliable measurement of the output state of the qubits is required for reading off the result of the computation and presently represents a formidable experimental challenge.

**Table 1.** Time scales of the qubit dynamics for the original [30] and improved [28] versions of the nuclear spin quantum computer with interactions mediated by the two-dimensional electron gas.

|  | The original model | The improved model |
|---|---|---|
| $T_{ext}$ | $O(10^{-5})$ sec | $O(10^{-5})$ sec |
| $T_{int}$ | $O(1)$ sec | $O(10^{-2})$ sec |
| $T_1$ | $O(10^3)$ sec | $O(10)$ sec |
| $T_2$ | $O(10)$ sec | $O(10^{-1})$ sec |